\pdfoutput=1

\documentclass[11pt,twoside,a4paper,cmspaper,final,collab]{cms-tdr}
\def\svnVersion{215780}\def\cmsCernNo{2012-292}\def\cmsCernDate{\today}\def\cmsMessage{Published in Physics Letters B as \href{http://dx.doi.org/10.1016/j.physletb.2013.09.009}{\doi{10.1016/j.physletb.2013.09.009.}}}

\begin{document}\cmsNoteHeader{EXO-12-012}

\hyphenation{had-ron-i-za-tion}
\hyphenation{cal-or-i-me-ter}
\hyphenation{de-vices}

\RCS$Revision: 146700 $
\RCS$HeadURL: svn+ssh://pakhotin@svn.cern.ch/reps/tdr2/papers/EXO-12-012/trunk/EXO-12-012.tex $
\RCS$Id: EXO-12-012.tex 146700 2012-09-10 14:21:29Z pakhotin $
\newlength\cmsFigWidth
\ifthenelse{\boolean{cms@external}}{\setlength\cmsFigWidth{0.85\columnwidth}}{\setlength\cmsFigWidth{0.4\textwidth}}
\ifthenelse{\boolean{cms@external}}{\providecommand{\cmsLeft}{top}}{\providecommand{\cmsLeft}{left}}
\ifthenelse{\boolean{cms@external}}{\providecommand{\cmsRight}{bottom}}{\providecommand{\cmsRight}{right}}
\newlength\myFigWidth
\setlength\myFigWidth{0.7\columnwidth}
\providecommand{\GeVccns}{\GeVcc}
\hyphenation{off-line}
\cmsNoteHeader{EXO-12-012} 
\title{Search for a non-standard-model Higgs boson decaying to a pair of new light bosons in four-muon final states}

\date{\today}

\abstract{Results are reported from a search for non-standard-model Higgs boson decays to pairs of new light bosons, each of which decays into the $\mu^+ \mu^-$ final state. The new bosons may be produced either promptly or via a decay chain. The data set corresponds to an integrated luminosity of 5.3\fbinv of proton-proton collisions at $\sqrt{s} = 7\TeV$, recorded by the CMS experiment at the LHC in 2011. Such Higgs boson decays are predicted in several scenarios of new physics, including supersymmetric models with extended Higgs sectors or hidden valleys. Thus, the results of the search are relevant for establishing whether the new particle observed in Higgs boson searches at the LHC has the properties expected for a standard model Higgs boson. No excess of events is observed with respect to the yields expected from standard model processes. A model-independent upper limit of $0.86 \pm 0.06$\unit{fb} on the product of the cross section times branching fraction times acceptance is obtained. The results, which are applicable to a broad spectrum of new physics scenarios, are compared with the predictions of two benchmark models as functions of a Higgs boson mass larger than 86\GeVcc and of a new light boson mass within the range 0.25--3.55\GeVcc.}

\hypersetup{%
pdfauthor={CMS Collaboration},%
pdftitle={Search for a non-standard-model Higgs boson decaying to a pair of new light bosons in four-muon final states},%
pdfsubject={CMS},%
pdfkeywords={LHC, CMS, Higgs, Supersymmetry, NMSSM, muons}}

\maketitle 

\section{Introduction \label{sec:introduction}}
The observation of a new particle~\cite{:2012gk, :2012gu} with a mass near $125\GeVcc$ in searches for the standard model (SM) Higgs boson~\cite{Englert:1964et, Higgs:1964pj, Guralnik:1964eu} at the Large Hadron Collider (LHC) raises the critical question of whether the new particle is in fact the SM Higgs boson. The precision of the comparisons of the new particle's production and decay properties with the final states predicted by the SM will improve with additional data. However, distinguishing a true SM Higgs boson from a non-SM Higgs bosons with couplings moderately different from the SM values will remain a challenge. Searches for non-SM Higgs boson production and decay modes are therefore particularly timely as they provide a complementary path, which in many cases can allow a discovery or rule out broad ranges of new physics scenarios with existing data.

This Letter presents a search for the production of a non-SM Higgs boson ($\text{h}$) decaying into a pair of new light bosons ($\text{a}$) of the same mass, which subsequently decay to pairs of oppositely charged muons (\textit{dimuons}) isolated from the rest of the event activity: $\text{h} \to 2\text{a} + \text{X} \to 4 \mu + \text{X}$, where $\text{X}$ denotes possible additional particles from cascade decays of a Higgs boson. This sequence of decays is predicted in several classes of models beyond the SM. One example is the next-to-minimal supersymmetric standard model (NMSSM)~\cite{Fayet1975104, Kaul198236, Barbieri1982343, Nilles1983346, Frere198311, Derendinger1984307, Drees:1988fc, Maniatis:2009re, Ellwanger:2009dp}, which extends the minimal supersymmetric standard model (MSSM)~\cite{Nilles:1983ge, Martin:1997ns, Chung:2003fi} by an additional gauge singlet field under new $U(1)_{PQ}$ symmetry in the Higgs sector of the superpotential. Compared to the MSSM, the NMSSM naturally generates the mass parameter $\mu$ in the Higgs superpotential at the electroweak scale~\cite{1984PhLB..138..150K} and significantly reduces the amount of fine tuning required~\cite{Casas:2003jx, Dermisek:2005ar, Chang:2008cw}. The Higgs sector of the NMSSM consists of $3$ $CP$-even Higgs bosons $\text{h}_{1,2,3}$ and $2$ $CP$-odd Higgs bosons $\text{a}_{1,2}$.

In the NMSSM, the $CP$-even Higgs bosons $\text{h}_1$ and $\text{h}_2$ can decay via $\text{h}_{1,2} \to 2\text{a}_1$, where one of the $\text{h}_1$ or $\text{h}_2$ is a SM-like Higgs boson that could correspond to the newly observed state at the LHC with a mass near $125\GeVcc$~\cite{:2012gk, :2012gu} and $\text{a}_1$ is a new $CP$-odd light Higgs boson~\cite{PhysRevD.39.844, Ellwanger:1993xa, Ellwanger:1995ru, Dobrescu:2000jt, Miller:2003ay}. The Higgs boson production cross section may differ substantially from that of the SM, depending on the parameters of a specific model. The new light boson $\text{a}_1$ couples weakly to SM particles, with the coupling to fermions proportional to the fermion mass, and can have a substantial branching fraction $\mathcal{B}(\text{a}_1 \to \mu^+ \mu^-)$ if its mass is within the range $2m_{\mu}<m_{\text{a}_1}<2m_{\tau}$~\cite{Belyaev:2010ka, Dermisek:2010mg}.

Pair production of light bosons can also occur in supersymmetric models with additional hidden (or \textit{dark}) valleys~\cite{ArkaniHamed:2008qn, Baumgart:2009tn, Falkowski:2010cm}, which are motivated by the excesses in positron spectra observed by satellite experiments~\cite{Adriani:2008zr, FermiLAT:2011ab}. These dark-SUSY models predict cold dark matter with a mass scale of ${\sim}1\TeVcc$, which can provide the right amount of relic density due to the Sommerfeld enhancement in the annihilation cross section arising from a new $U(1)_{D}$ symmetry~\cite{Hisano:2003ec, Cirelli:2008pk}. In these models, $U(1)_D$ is broken, giving rise to light but massive dark photons $\gamma_D$ that weakly couple to the SM particles via a small kinetic mixing~\cite{Holdom:1985ag, Dienes:1996zr, Cheung:2009qd} with photons. The lightest neutralino $\text{n}_1$ in the \textit{visible} (as opposed to \textit{hidden}) part of the SUSY spectrum is no longer stable and can decay via e.g. $\text{n}_1 \rightarrow \text{n}_D + \gamma_D$, where $\text{n}_D$ is a light dark fermion (dark neutralino) that escapes detection. The SM-like Higgs boson can decay via $\text{h} \to 2 \text{n}_1$, if $m_{\text{h}} > 2m_{\text{n}_1}$. The branching fraction $\mathcal{B}(\text{h} \to 2 \text{n}_1)$  can vary from very small to large, bounded by the LHC measurements in the context of Higgs searches, since the bounds obtained at LEP can be circumvented~\cite{Falkowski:2010cm}. The lack of an anti-proton excess in the measurements of the cosmic ray spectrum constrains the mass of $\gamma_D$ to be ${\leq}\mathcal{O}(1)\GeVcc$~\cite{Adriani:2010rc}. Assuming that $\gamma_D$ can only decay to SM particles, the branching fraction $\mathcal{B}(\gamma_D \to \mu^+ \mu^-)$ can be as large as $45\%$, depending on $m_{\gamma_D}$~\cite{Falkowski:2010cm}. The Higgs boson production cross section may or may not be enhanced compared to the SM, depending on the specific parameters of the model. The search described in this Letter was designed to be independent of the details of specific models, and the results can be interpreted in the context of other models predicting the production of the same final states.

Previous searches for the pair production of new light bosons decaying into dimuons were performed at the Tevatron with a 4.2\fbinv data sample~\cite{Abazov:2009yi} and more recently at the LHC with a 35\pbinv~\cite{Chatrchyan:2011hr} and a 1.9\fbinv~\cite{Aad:2012kw} data samples. Associated production of the light $CP$-odd scalar bosons has been searched for at $\Pep\Pem$ colliders~\cite{Love:2008aa,Aubert:2009cp} and the Tevatron~\cite{Aaltonen:2011aj}. Direct production of the $\text{a}_1$ has been studied at the LHC~\cite{Chatrchyan:2012am}, but in the framework of NMSSM the sensitivity of these searches is limited by the typically very weak coupling of the $\text{a}_1$ to SM particles. The most stringent limits on  the Higgs sector of the NMSSM are provided by the WMAP data~\cite{Jarosik:2010iu} and LEP searches~\cite{Abbiendi:2002qp, Abbiendi:2002in, Schael:2006cr} ($m_{\text{h}_1} > 86\GeVcc$). In the framework of dark SUSY, experimental searches for $\gamma_D$ have focused on the production of dark photons at the end of SUSY cascades at the Tevatron~\cite{Abazov:2009hn, Abazov:2010uc, Aaltonen:2012pu} and the LHC~\cite{Chatrchyan:2011hr}. Furthermore, if the newly observed particle at the LHC~\cite{:2012gk, :2012gu} is indeed a Higgs boson, the studies of its SM decays will provide additional constraints on the allowed branching fractions for the non-SM decays.

\section{The CMS detector \label{sec:detector}}
The analysis presented in this Letter uses experimental data collected by the Compact Muon Solenoid (CMS) experiment at the LHC in 2011. The central feature of the CMS apparatus is a superconducting solenoid of 6\unit{m} internal diameter, providing a magnetic field of 3.8\unit{T}. Within the superconducting solenoid volume are a silicon pixel and strip tracker, a lead tungstate crystal electromagnetic calorimeter, and a brass/scintillator hadron calorimeter. The inner tracker measures charged particles within the pseudorapidity range $\abs{\eta} < 2.5$, where $\eta = -\ln [\tan(\theta/2)]$ and $\theta$ is the polar angle with respect to the direction of the counterclockwise proton beam that is the $z$-axis of the CMS reference frame. The tracker provides an impact parameter resolution of ${\sim}15\mum$ and a transverse momentum ($\pt$) resolution of about $1.5\%$ for 100\GeVc particles. Muons are measured in gas-ionization detectors embedded in the steel return yoke. The muon detectors are made using the following technologies: drift tubes ($\abs{\eta} < 1.2$), cathode strip chambers ($0.9 < \abs{\eta} < 2.4$), and resistive-plate chambers ($\abs{\eta} < 1.6$). Matching the muons to the tracks measured in the silicon tracker results in a transverse momentum resolution between 1 and 5\% for $\pt$ values up to 1\TeVc. A more detailed description can be found in Ref.~\cite{Chatrchyan:2008aa}.

\section{Data selection \label{sec:selection}}
The search is performed as a ``blind'' analysis, \ie data in the signal region were not used to define the reconstruction and selection procedures. The analysis is based on a data sample corresponding to an integrated luminosity of 5.3\fbinv of proton-proton collisions at $\sqrt{s} = 7\TeV$, obtained in 2011. The data were collected with a trigger selecting events containing at least two muons, one with $\pt > 17\GeVc$ and one with $\pt > 8\GeVc$. In the offline analysis, events are selected by requiring at least one primary vertex reconstructed with at least four tracks and with its $z$ coordinate within 24\unit{cm} of the nominal collision point. Offline muon candidates are built using tracks reconstructed in the inner tracker matched to track segments in the muon system, using an arbitration algorithm~\cite{CMS_Muon_Reco}. The candidates are further required to have at least eight hits in the tracker, with the $\chi^2/\text{Ndof} < 4$ for the track fit in the inner tracker (where $\text{Ndof}$ is the number of degrees of freedom), and at least two matched segments in the muon system. The data are further selected by requiring at least four offline muon candidates with $\pt > 8\GeVc$ and $\abs{\eta} < 2.4$;  at least one of the candidates must have $\pt > 17\GeVc$ and be reconstructed in the central region, $\abs{\eta} < 0.9$. Application of the selection requirements described above yields $1,745$ events in the data. The trigger efficiency for the selected events is high (96--97\%) and is nearly independent of the $\pt$ and $\eta$ of any of the four muons. The $\abs{\eta}<0.9$ requirement is tighter than that imposed by the trigger, but eliminates significant model dependence attributable to the reduced trigger performance in the forward region in the presence of multiple spatially close muons. This $\eta$ requirement causes an overall reduction in the analysis acceptance of about 20\%, as obtained in a simulation study with one of the NMSSM benchmark samples used in the analysis.

Next, oppositely charged muons are grouped into dimuons (a muon may be shared between several dimuons) if their pairwise invariant mass satisfies $m_{\mu\mu} < 5\GeVcc$ and if either the fit of the two muon tracks for a common vertex has a $\chi^2$ fit probability greater than $1$\% or the two muon tracks satisfy the cone size requirement $\Delta R (\mu^+, \mu^-) = \sqrt{(\eta_{\mu^+} - \eta_{\mu^-})^2 + (\phi_{\mu^+} - \phi_{\mu^-})^2} < 0.01$, where $\phi$ is the azimuthal angle in radians. The $\Delta R$ requirement compensates for the reduced efficiency of the vertex probability requirement for dimuons with very low mass ($m_{\mu\mu} \gtrsim 2 m_{\mu}$), in which the two muon tracks are nearly parallel to each other at the point of closest approach.

Once all dimuons are constructed, only events with exactly two dimuons not sharing common muons are selected for further analysis. There is no restriction on the number of ungrouped \textit{(orphan)} muons. Assuming that each dimuon is a decay product of a new light boson, we require that the two dimuons have invariant masses in the range 0.25--3.55\GeVcc. We reconstruct $z_{\mu\mu}$, the projected $z$ coordinate of the dimuon system at the point of the closest approach to the beam line, using the dimuon momentum measured at the common vertex and the vertex position. We ensure that the two dimuons originate from the same $\text{pp}$ interaction by requiring $|z_{\mu\mu_1} - z_{\mu\mu_2}|<1$\unit{mm}. This selection yields $139$ events in data and it is fully efficient for signal events while reducing the probability of selecting rare events with dimuons from two separate primary interactions.

To suppress backgrounds with dimuons coming from jets, we require that the dimuons be isolated from other activity in the event, using the criterion $I_{\text{sum}} < 3\GeVc$, where the isolation parameter of the dimuon system $I_{\text{sum}}$ is defined as the scalar sum of the transverse momenta of all additional charged tracks with $\pt > 0.5\GeVc$  within a cone of size $\Delta R=0.4$ centered on the momentum vector of the dimuon system. Tracks used in the calculation of $I_{\text{sum}}$ must also have a $z$ coordinate at the point of the closest approach to the beam line that lies within 1\unit{mm} of the $z$ coordinate of the dimuon system. The $I_{\text{sum}}$ selection yields three events in data and it suppresses the contamination from $\bbbar$ production by about a factor of 40 (measured in data) while rejecting less than 10\% of the signal events (obtained from the simulation study).

Finally, we require that the invariant masses of the two reconstructed dimuons are compatible with each other within the detector resolution $|m_1 - m_2| < 0.13\GeVcc + 0.065 \times (m_1+m_2)/2$, where $m_1 = m_{\mu\mu_1}$ and $m_2 = m_{\mu\mu_2}$. The numerical parameters in this last requirement correspond to at least five times the size of the core resolution in dimuon mass, including the differences in resolution in the central and forward regions. The signal inefficiency of this $m_1 \simeq m_2$ selection is less than 5\% per event; it is due to QED final-state radiation and is unrelated to the detector resolution. No constraint is imposed on the four-muon invariant mass, in order to maintain the model independence of the analysis, in particular with respect to models resulting in cascade decays such as dark-SUSY, where an unknown fraction of the energy goes into the light dark fermions, which escape detection.

\begin{table*}[t]
\topcaption{Event selection efficiencies $\epsilon^\mathrm{MC}_{\text{full}}(m_{\text{h}_1}, m_{\text{a}_1})$, as obtained from the full detector simulation, and the geometric and kinematic acceptances $\alpha_{\text{gen}}(m_{\text{h}_1}, m_{\text{a}_1})$ calculated using generator level information only, with statistical uncertainties for the NMSSM benchmark model. The experimental data-to-simulation scale factors are not applied.\label{tab:efficiency_NMSSM}}
\begin{center}
\begin{footnotesize}
\begin{tabular}{l c c c c c c c c}
\hline
\hline
$m_{\text{h}_1} \: [\GeVccns{}]$	&	$	90			$	&	$	100			$	&	$	125			$	&	$	125			$	&	$	125			$	&	$	125			$	&	$	125			$	&	$	150			$	\\
$m_{\text{a}_1} \: [\GeVccns{}]$	&	$	2			$	&	$	2			$	&	$	0.25			$	&	$	0.5			$	&	$	1			$	&	$	2			$	&	$	3			$	&	$	2			$	\\
$\epsilon^\mathrm{MC}_{\text{full}} \: [\%]$	&	$	12.1	\pm	0.1	$	&	$	14.7	\pm	0.1	$	&	$	46.2	\pm	0.1	$	&	$	24.6	\pm	0.2	$	&	$	21.1	\pm	0.1	$	&	$	20.1	\pm	0.1	$	&	$	19.7	\pm	0.1	$	&	$	24.0	\pm	0.1	$	\\
$\alpha_{\text{gen}} \: [\%]$	&	$	16.6	\pm	0.1	$	&	$	20.0	\pm	0.1	$	&	$	62.2	\pm	0.1	$	&	$	33.2	\pm	0.3	$	&	$	28.6	\pm	0.1	$	&	$	27.5	\pm	0.1	$	&	$	27.1	\pm	0.1	$	&	$	33.2	\pm	0.1	$	\\
$\epsilon^\mathrm{MC}_{\text{full}}/\alpha_{\text{gen}} \: [\%]$	&	$	73.0	\pm	0.3	$	&	$	73.5	\pm	0.3	$	&	$	74.3	\pm	0.3	$	&	$	74.2	\pm	0.6	$	&	$	73.8	\pm	0.3	$	&	$	72.6	\pm	0.3	$	&	$	72.7	\pm	0.3	$	&	$	72.2	\pm	0.2	$	\\
\hline
\hline
\end{tabular}
\end{footnotesize}
\end{center}
\end{table*}

To demonstrate the ability of the analysis to select a possible signal, we use the two benchmark models introduced earlier. The NMSSM samples are simulated with the \PYTHIA~6.4.26 event generator~\cite{Sjostrand:2006za} using MSSM Higgs boson production via gluon-gluon fusion $\text{g}\text{g} \to \text{H}^0_\mathrm{MSSM}$, where the Higgs bosons are forced to decay via $\text{H}^0_\mathrm{MSSM} \to 2\text{A}^0_\mathrm{MSSM}$. The masses of $\text{H}^0_\mathrm{MSSM}$ and $\text{A}^0_\mathrm{MSSM}$ are set to the desired values for the $\text{h}_1$ mass and $\text{a}_1$ mass, respectively. Both $\text{A}^0_\mathrm{MSSM}$ bosons are forced to decay to a pair of muons. The dark-SUSY samples are simulated with the \MADGRAPH~4.5.2 event generator~\cite{Alwall:2007st} using SM Higgs boson production via gluon-gluon fusion $\text{g}\text{g} \to \text{h}_\mathrm{SM}$, where the mass of $\text{h}_\mathrm{SM}$ is set to the desired value for the $\text{h}$ mass. The \textsc{bridge} software~\cite{Meade:2007js} was used to implement the new physics model that forces the Higgs bosons $\text{h}_\mathrm{SM}$ to undergo a non-SM decay to a pair of neutralinos $\text{n}_1$, each of which decays $\text{n}_1 \to \text{n}_D + \gamma_D$, where $m_{\text{n}_1} = 10\GeVcc$, $m_{\text{n}_D} = 1\GeVcc$ and $m_{\gamma_D} = 0.4\GeVcc$. Both dark photons $\gamma_D$ are forced to decay to two muons, while both dark neutralinos $\text{n}_D$ escape detection. The narrow width approximation is imposed by setting the widths of the Higgs bosons and dark photons to a small value ($10^{-3}\GeVcc$). All benchmark samples are generated using the leading-order CTEQ6L1~\cite{Pumplin:2002vw} set of parton distribution functions (PDF), and are interfaced with \PYTHIA~6.4.26 using the Z2 tune~\cite{Chatrchyan:2011id} for ``underlying event'' (UE) activity at the LHC and to simulate jet fragmentation, when applicable.

All events in the benchmark signal samples are processed through a detailed simulation of the CMS detector based on \GEANTfour~\cite{Allison:2006ve} and are reconstructed with the same algorithms used for data analysis. Tables~\ref{tab:efficiency_NMSSM} and~\ref{tab:efficiency_SUSY} show the event selection efficiencies $\epsilon^\mathrm{MC}_{\text{full}}$ obtained using the simulated signal events for these two benchmark models using representative choices for masses of $\text{h}$, $\text{a}_1$ or $\gamma_D$. To provide a simple recipe for future reinterpretations of the results in the context of other models, we separately determine $\alpha_{\text{gen}}$, the geometric and kinematic acceptance of this analysis calculated using generator level information only. It is defined with the criteria that an event contains at least four muons with $\pt > 8\GeVc$ and $\abs{\eta} < 2.4$, with at least one of these muons having $\pt > 17\GeVc$ and $\abs{\eta} < 0.9$. Tables~\ref{tab:efficiency_NMSSM} and \ref{tab:efficiency_SUSY} also show $\alpha_{\text{gen}}$ along with the ratio $\epsilon^\mathrm{MC}_{\text{full}}/\alpha_{\text{gen}}$.

\begin{table}[t]
\topcaption{Event selection efficiencies $\epsilon^\mathrm{MC}_{\text{full}}(m_{\text{h}}, m_{\gamma_D})$, as obtained from the full detector simulation, and the geometric and kinematic acceptances $\alpha_{\text{gen}}(m_{\text{h}}, m_{\gamma_D})$ calculated using generator level information only, with statistical uncertainties for a dark-SUSY benchmark model, as obtained from simulation. The experimental data-to-simulation scale factors are not applied.\label{tab:efficiency_SUSY}}
\begin{center}
\begin{footnotesize}
\begin{tabular}{l c c c}
\hline
\hline
$m_{\text{h}} \: [\GeVccns{}]$	&	$	90			$	&	$	125			$	&	$	150			$	\\
$m_{\gamma_D} \: [\GeVccns{}]$	&	$	0.4			$	&	$	0.4			$	&	$	0.4			$	\\
$\epsilon^\mathrm{MC}_{\text{full}} \: [\%]$	&	$	2.7	\pm	0.1	$	&	$	7.6	\pm	0.1	$	&	$	11.4	\pm	0.1	$	\\
$\alpha_{\text{gen}} \: [\%]$	&	$	3.6	\pm	0.1	$	&	$	10.1	\pm	0.1	$	&	$	15.2	\pm	0.1	$	\\
$\epsilon^\mathrm{MC}_{\text{full}}/\alpha_{\text{gen}} \: [\%]$	&	$	76.1	\pm	0.8	$	&	$	75.5	\pm	0.5	$	&	$	74.9	\pm	0.4	$	\\
\hline
\hline
\end{tabular}
\end{footnotesize}
\end{center}
\end{table}

The model independence of the ratio $\epsilon^\mathrm{MC}_{\text{full}}/\alpha_{\text{gen}}$ permits an estimate of the full event selection efficiency of this analysis for an arbitrary new physics model predicting the signature with a pair of new light bosons. The analysis makes some assumptions on the nature and characteristics of the new light bosons, namely that they should be of the same type, have a mass in range $2m_{\mu} < m_\text{a} < 2 m_{\tau}$, decay to the $\mu^+ \mu^-$ final state and not have a significant lifetime. In addition, the two bosons, should be isolated and sufficiently separated from each other to avoid being vetoed by the isolation requirement. The acceptance $\alpha_{\text{gen}}$ may be calculated using only the generator level selection requirements that have been defined above. The full efficiency $\epsilon_{\text{full}}$ could then be calculated by multiplying $\alpha_{\text{gen}}$ by the ratio $\epsilon_{\text{full}}/\alpha_{\text{gen}} = r \times \epsilon^\mathrm{MC}_{\text{full}}/\alpha_{\text{gen}} = 0.67 \pm 0.05$, where $r = \epsilon_{\text{full}}/\epsilon^\mathrm{MC}_{\text{full}} = 0.91 \pm 0.07$ is the scale factor defined in Sec.~\ref{sec:systematics} that accounts for differences between data and simulation, and $\epsilon^\mathrm{MC}_{\text{full}}/\alpha_{\text{gen}} = 0.74 \pm 0.02$ is an average ratio over all of the benchmark points used. The systematic uncertainty in the ratio $\epsilon_{\text{full}}/\alpha_{\text{gen}}$ is around 7.4\%. For reference, the individual systematics uncertainties in $\alpha_{\text{gen}}$ and $\epsilon_{\text{full}}$ are 3.0\% and 8.0\%, respectively, as discussed in Sec.~\ref{sec:systematics}.

\section{Background estimation \label{sec:background}}
The background contributions after final selections are dominated by $\bbbar$ and direct $\JPsi$ pair production events. The leading part of the $\bbbar$ contribution is due to $\cPqb$-quark decays to pairs of muons via double semileptonic decays or resonances, i.e. $\omega$, $\rho$, $\phi$, $\JPsi$. A smaller contribution comes from events with one real dimuon and a second dimuon with a muon from a semileptonic \cPqb-quark decay and a charged hadron misidentified as another muon. The misidentification typically occurs due to the incorrect association of the track of the charged hadron with the track segments from a real muon in the muon system. The contribution of other SM processes has been found to be negligible (less than $0.1$ events combined), for example low mass Drell--Yan production is heavily suppressed by the requirement of additional muons, and $\text{p} \text{p} \to \text{Z}/\gamma^* \to 4 \mu$ production is suppressed by the requirement of small and mutually consistent masses of the dimuons~\cite{Belyaev:2010ka}. The analysis is not sensitive to SM process $\text{p} \text{p} \to \text{H} \to \text{ZZ} \to 4 \mu$ because the invariant mass of the dimuons is substantially lower than the $\text{Z}$ mass.

\begin{figure*}[t]
\includegraphics[width=0.4642\textwidth]{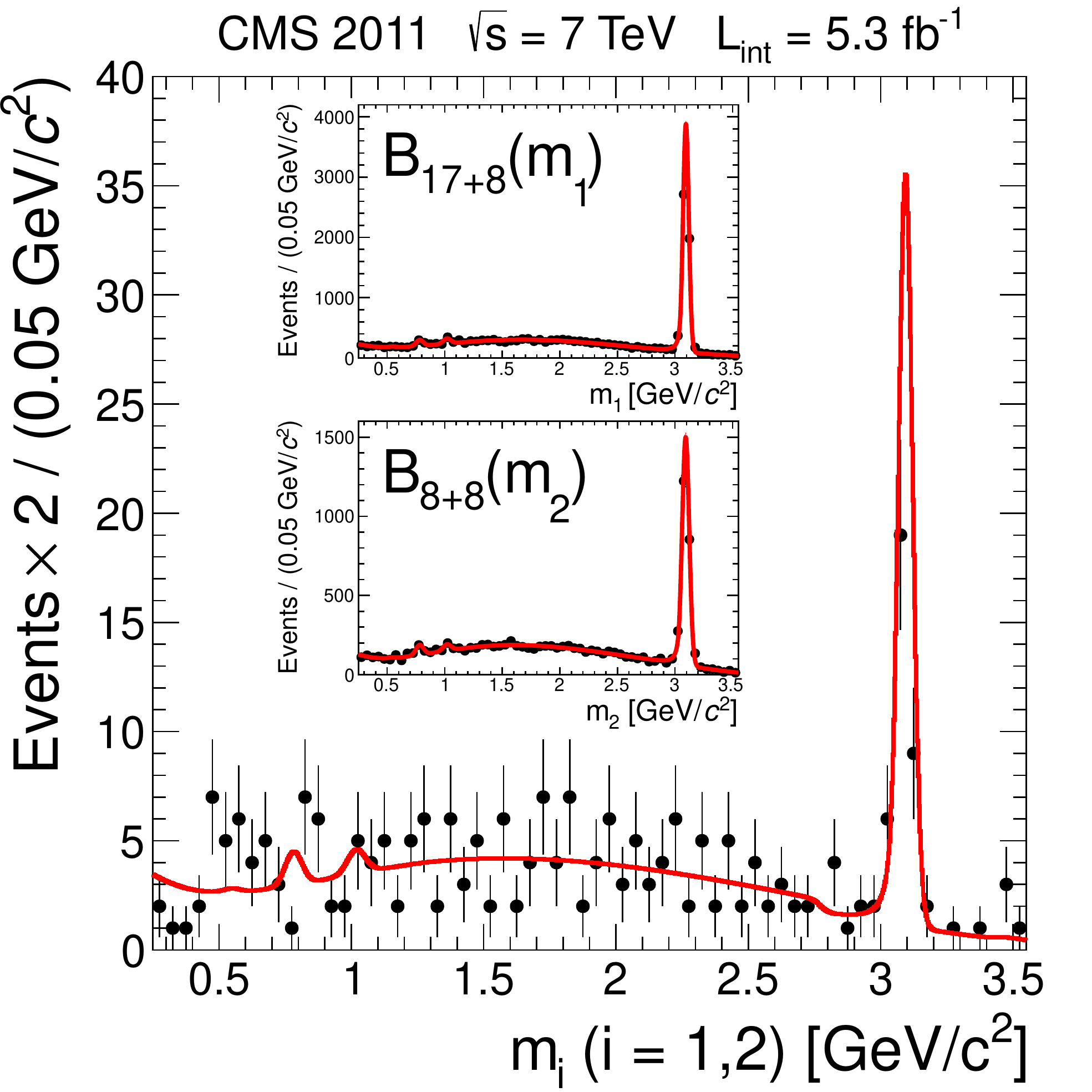} 
\hfill
\includegraphics[width=0.5158\textwidth]{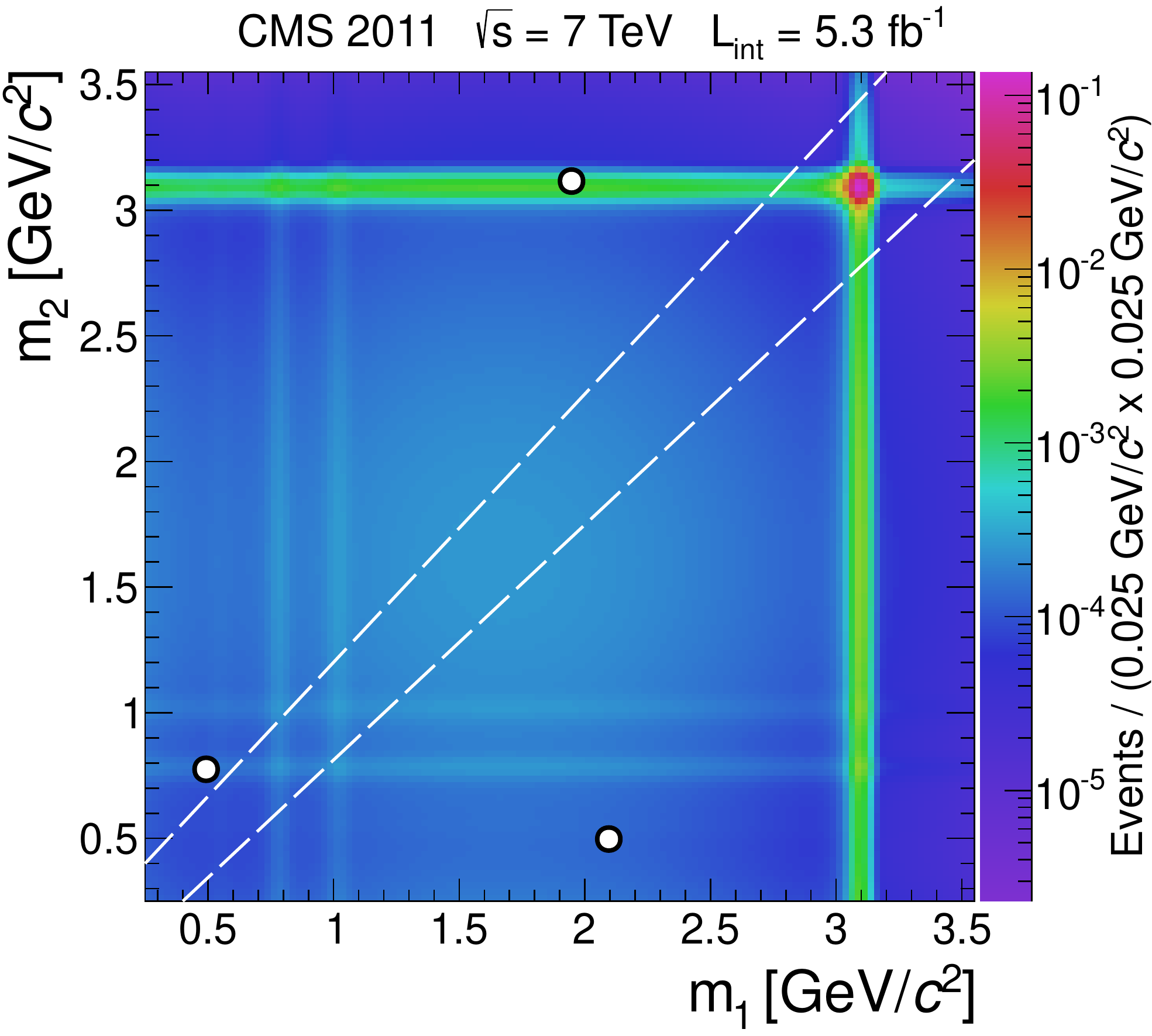} 
\caption{Left: Comparison of the data (solid circles) failing the $m_1 \simeq m_2$ requirement in the control sample where no isolation requirement is applied to reconstructed dimuons with the prediction of the background shape model (solid line) scaled to the number of entries in the data. The insets show the $B_{17+8}$ and $B_{8+8}$ templates (solid lines) for dimuons obtained with background-enriched data samples. Right: Distribution of the invariant masses $m_1$ vs. $m_2$ for the isolated dimuon systems for the three events in the data (shown as empty circles) surviving all selections except the requirement that these two masses fall into the diagonal signal region $m_1 \simeq m_2$ (outlined with dashed lines). The intensity (color online) of the shading indicates the background expectation which is a sum of the $\bbbar$ and the direct $\JPsi$ pair production contributions.
\label{fig:control}}
\end{figure*}

Using data control samples, the $\bbbar$ background is modeled as a two dimensional (2D) template $B_{\bbbar}(m_1,m_2)$ in the plane of the invariant masses of the two dimuons in the selected events, where $m_1$ always refers to the dimuon containing a muon with $\pt > 17\GeVc$ and $\abs{\eta}<0.9$. For events with both dimuons containing such a muon, the assignment of $m_1$ and $m_2$ is random. As each $\cPqb$ quark fragments independently, we construct the template describing the 2D probability density function as a Cartesian product $B_{17+8}(m_1) \times B_{8+8}(m_2)$, where the $B_{17+8}$ and $B_{8+8}$ templates model the invariant-mass distributions for dimuons with or without the requirement that the dimuon contains at least one muon satisfying $\pt > 17\GeVc$ and $\abs{\eta}<0.9$. This distinction is necessary as the shape of the dimuon invariant mass distribution depends on the transverse momentum thresholds used to select muons and whether the muons are in the central ($\abs{\eta}<0.9$) or in the forward ($0.9<\abs{\eta}<2.4$) regions, owing to the differences in momentum resolution of the barrel and endcap regions of the tracker. The $B_{17+8}$ shape is measured using a data sample enriched in $\bbbar$ events with exactly one dimuon and one orphan muon under the assumption that one of the $\cPqb$ quarks decays to a dimuon containing at least one muon with $\pt > 17\GeVc$ and $\abs{\eta}<0.9$, while the other $\cPqb$ quark decays semileptonically resulting in an orphan muon with $\pt > 8\GeVc$. For the $B_{8+8}$ shape, we use a similar sample and procedure but only require the dimuon to have both muons with $\pt > 8\GeVc$, while the orphan muon has to have $\pt > 17\GeVc$ and $\abs{\eta}<0.9$. Both data samples used to measure background shapes are collected with the same trigger and with kinematic properties similar to those $\bbbar$ events passing the selections of the main analysis. These event samples do not overlap the sample containing two dimuons that is  used for the main analysis, and they have negligible contributions from non-$\bbbar$ backgrounds. The $B_{17+8}$ and $B_{8+8}$ distributions, fitted with a parametric analytical function using a combination of Bernstein polynomials~\cite{Bernstein} and Crystal Ball functions~\cite{Oreglia:1980cs} describing resonances, are shown as insets in Fig.~\ref{fig:control}~(left). Once the $B_{\bbbar}(m_1,m_2)$ template is constructed, it is used to provide a description of the $\bbbar$ background shape in the main analysis.

To validate the constructed $B_{\bbbar}(m_1,m_2)$ template, we compare its shape with the distribution of the invariant masses $m_1$ vs. $m_2$ from events obtained with all standard selections except the requirement that each of the two reconstructed dimuons is isolated. Omitting the isolation requirement provides a high-statistics control sample of events with two dimuons highly enriched with $\bbbar$ events. To avoid unblinding the search, the diagonal signal region is excluded in both the data and the template, i.e. the comparison has been limited to the data events that satisfy all analysis selections but fail the $m_1 \simeq m_2$ requirement. Distributions of $m_1$ and $m_2$ are consistent with the projections of the $B_{\bbbar}(m_1,m_2)$ template on the respective axes normalized to the number of events in the data control sample. The sum of the $m_1$ and $m_2$ distributions agrees well with the sum of the template projections as shown in Fig.~\ref{fig:control}~(left).

Another cross-check has been performed using data events which satisfy all analysis selections except that the isolation parameters of each dimuon system have been required to satisfy $3\GeVc < I_{\text{sum}} < 8\GeVc$, which removes potential signal events since the signal selections require $I_{\text{sum}}<3\GeVc$ for each dimuon. These selections with both dimuons in isolation sideband yield four events in the off-diagonal region and zero events in the diagonal region of $(m_1,m_2)$ plane. Normalizing the background distribution to these four observed events, we predict $0.9 \pm 0.4$ $\bbbar$ events in the diagonal region of the $(m_1,m_2)$ plane with both dimuons in the isolation sideband ($3\GeVc < I_{\text{sum}} < 8\GeVc$). This prediction is consistent with no events being observed there.

To normalize the constructed $B_{\bbbar}(m_1,m_2)$ template, we use the data events that satisfy all analysis selections, but fail the $m_1 \simeq m_2$ requirement. These selections yield three events in the off-diagonal sideband region of the $(m_1,m_2)$ plane, leading, in the diagonal signal region, to an expected number of $0.7 \pm 0.4$ $\bbbar$ events, where the estimated uncertainty is dominated by the statistical uncertainty. These three events in the off-diagonal sidebands of the $(m_1,m_2)$ plane are shown as empty circles in Fig.~\ref{fig:control}~(right).

The direct \JPsi pair production contribution is estimated using simulations normalized to the data in the region of low invariant mass of the two \JPsi candidates. The data for the study were collected using a trigger requiring three muon candidates with transverse momenta $\pt > 1\GeVc$ and a scalar sum of momenta $p (\mu_1)+p(\mu_2)+p(\mu_3) > 2.5\GeVc$. In addition, among these muon candidates there must be at least one pair with opposite charges, originating from a common vertex and having an invariant mass in the range $2.8 < m_{\mu\mu} < 3.35\GeVcc$. In the offline selection, four high quality muon candidates are required, forming two \JPsi candidates within this same mass window. The resulting data sample has significant contamination from non-prompt \JPsi pairs produced in heavy flavor decays, combinatorial backgrounds, and combinations of the two. The rate of prompt double \JPsi production in the data sample is extracted by extrapolating in the plane of the transverse lifetime $ct_{xy}$ of the two \JPsi candidates and their measured invariant masses. The rate thus obtained from data is used to normalize the simulation, which is produced using the double parton scattering (DPS) \JPsi pair production process in the \PYTHIA~8.108 event generator~\cite{Sjostrand:2007gs}, applying the same selections as applied to the data. This calculation provides an estimate of the background rate due to prompt double \JPsi production in the signal region. To evaluate the systematic uncertainty, this normalization factor is recalculated separately for each of the four ranges of the invariant masses $m_{2\JPsi}$ of the two \JPsi candidates: 6--13, 13--22, 22--35, and 35--80\GeVcc. Although the control samples overlap with the data sample used for the search, for none of the benchmark models considered are the control samples significantly contaminated with the signal. This is because the dark-SUSY models predict new bosons with a mass of less than $1\GeVcc$, and LEP measurements set an upper limit of about $90\GeVcc$ on the NMSSM CP-even Higgs boson mass. With approximately 1500 events in the prompt double \JPsi control sample, the statistical component of the prompt double \JPsi background uncertainty is very small compared to the systematic error. For this reason, the correlation introduced into the limit calculation by the potential presence of a small amount of signal in the control region may be safely neglected. Nevertheless, we have treated the higher mass ranges with particular caution, as an excess there could be a sign of a potential signal of new physics, in which case it would have been necessary to change the strategy of this analysis. Measurements in all ranges of  $m_{2\JPsi}$ yield consistently low estimates of the contamination due to the prompt double \JPsi background in the signal region. The final estimate of the rate is $0.3 \pm 0.3$, where the uncertainty accounts for both statistical and systematic effects. This value is used to normalize a 2D Gaussian template $B_{2 \JPsi}(m_1, m_2)$ in the $(m_1, m_2)$ plane that models the double \JPsi background contribution.

The distribution of the total background expectation in the $(m_1,m_2)$ plane is $B_{\bbbar}(m_1,m_2) + B_{2 \JPsi}(m_1, m_2)$, \ie a sum of the $\bbbar$ and the direct $\JPsi$ pair production contributions. It is shown by the intensity of the shading in Fig.~\ref{fig:control}~(right). The background expectation in the diagonal signal region is $1.0 \pm 0.5$ events, where the uncertainty accounts for both statistical and systematic effects.

\begin{figure*}[t]
\begin{center}
\includegraphics[width=0.49\textwidth]{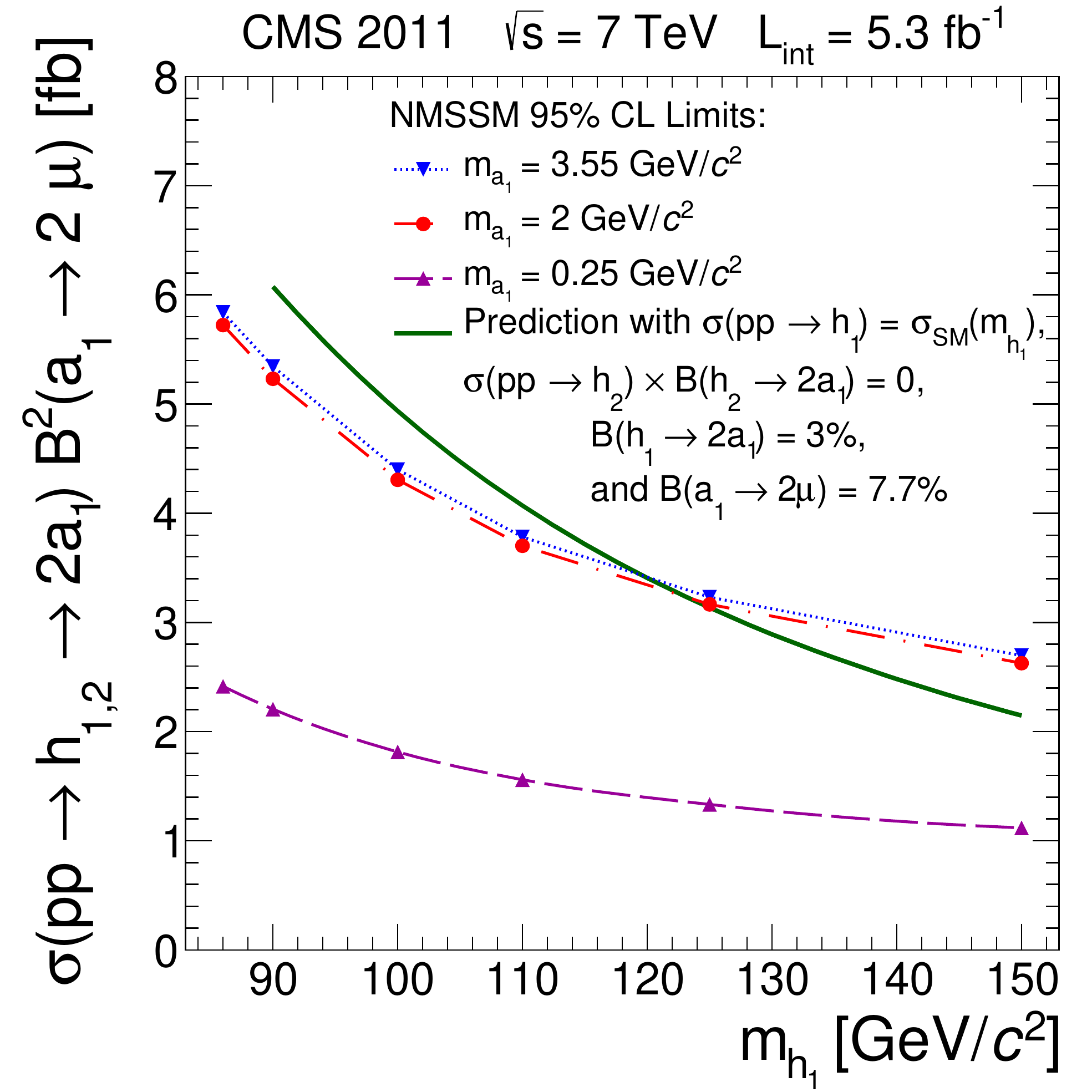}
\hfill
\includegraphics[width=0.49\textwidth]{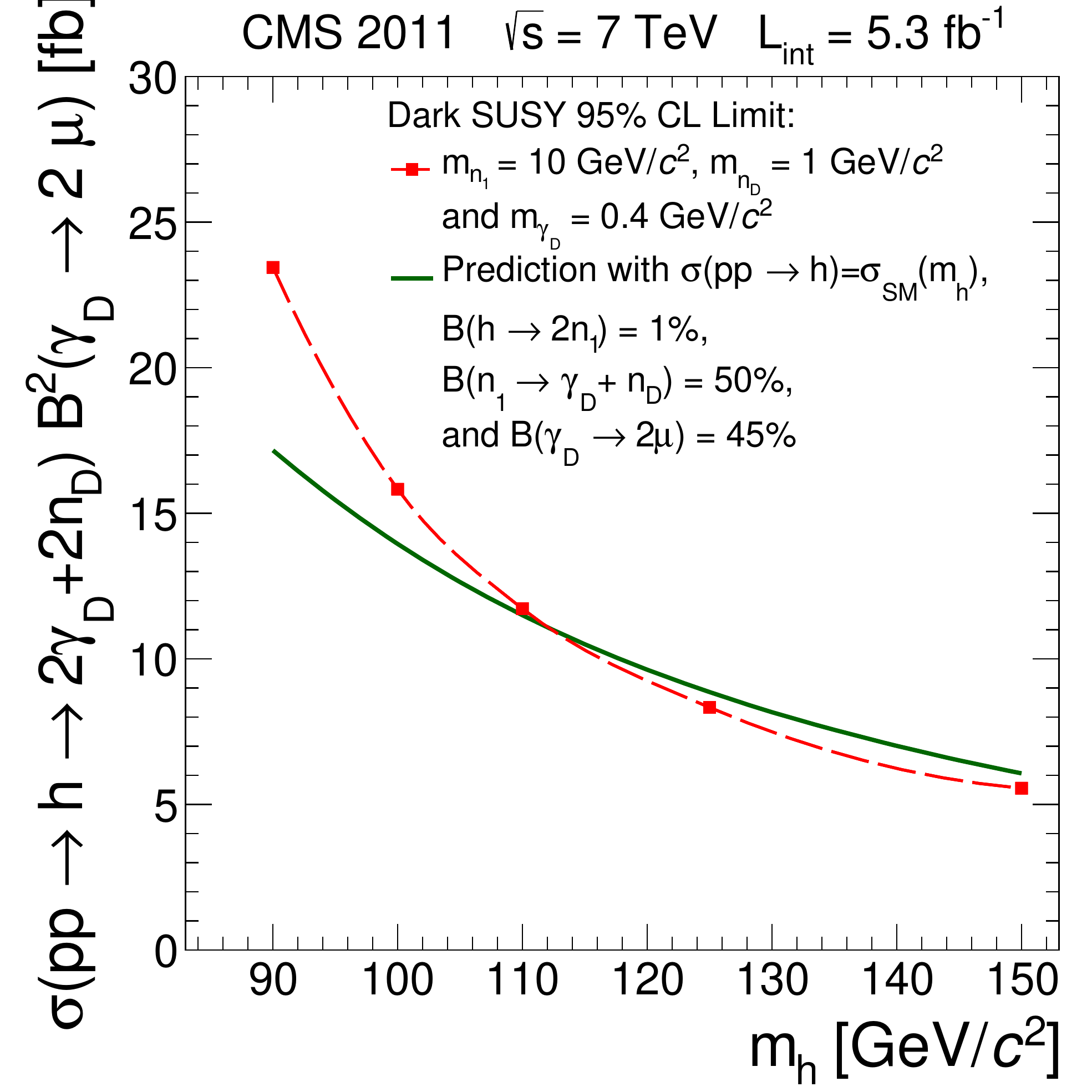}
\end{center}
\caption{Left: The $95\%$ CL upper limits as functions of $m_{\text{h}_1}$, for the NMSSM case, on $\sigma(\Pp\Pp \to \text{h}_{1,2} \to 2 \text{a}_1) \times \mathcal{B}^2(\text{a}_1 \to 2 \mu)$ with $m_{\text{a}_1}=0.25\GeVcc$ (dashed curve), $m_{\text{a}_1}=2\GeVcc$ (dash-dotted curve) and $m_{\text{a}_1}=3.55\GeVcc$ (dotted curve). As an illustration, the limits are compared to the predicted rate (solid curve) obtained using a simplified scenario with $\sigma (\Pp\Pp \to \text{h}_1)=\sigma_\mathrm{SM}(m_{\text{h}_1})$~\cite{Dittmaier:2011ti}, $\sigma (\Pp\Pp \to \text{h}_2) \times \mathcal{B}(\text{h}_{2} \rightarrow 2 \text{a}_1) = 0$, $\mathcal{B}(\text{h}_1 \to 2 \text{a}_1) = 3\%$, and $\mathcal{B}(\text{a}_1 \to 2\mu) = 7.7\%$. The chosen $\mathcal{B}(\text{a}_1 \to 2\mu)$ is taken from~\cite{Dermisek:2010mg} for $m_{\text{a}_1} = 2\GeVcc$ and NMSSM parameter $\tan \beta = 20$. Right: The 95\% CL upper limit as a function of $m_{\text{h}}$, for the dark-SUSY case, on $\sigma(\Pp\Pp \to \text{h} \to 2 \text{n}_1 \to 2\text{n}_D + 2 \gamma_D) \times \mathcal{B}^2(\gamma_D \to 2 \mu)$ with $m_{\text{n}_1}=10\GeVcc$, $m_{\text{n}_D}=1\GeVcc$ and $m_{\gamma_D}=0.4\GeVcc$ (dashed curve). As an illustration, the limit is compared to the predicted rate (solid curve) obtained using a simplified scenario with SM Higgs boson production cross section $\sigma (\Pp\Pp \to \text{h})=\sigma_\mathrm{SM}(m_{\text{h}})$~\cite{Dittmaier:2011ti}, $\mathcal{B}(\text{h} \to 2 \text{n}_1)=1\%$, $\mathcal{B}(\text{n}_1 \to \text{n}_D + \gamma_D)=50\%$, and $\mathcal{B}(\gamma_D \to 2 \mu) = 45\%$. The chosen $\mathcal{B}(\gamma_D \to 2 \mu)$ is taken from~\cite{Falkowski:2010cm} for $m_{\gamma_D} = 0.4\GeVcc$.
\label{fig:results}}
\end{figure*}

\section{Systematic uncertainties \label{sec:systematics}}
The selection efficiencies of offline muon reconstruction, trigger, and dimuon isolation criteria are obtained with simulation and have been corrected with scale factors derived from a comparison of data and simulation using $\cPZ \to \mu \mu$ and $\JPsi \to \mu \mu$ samples. The scale factor per event is $r = 0.91 \pm 0.07\syst.$ It accounts for the differences in the efficiency of the trigger, the efficiency of the muon reconstruction and identification for each of the four muon candidates, and the combined efficiency of the isolation requirement for the two dimuon candidates. The correlations due to the presence of two close muons have been taken into account. The main systematic uncertainty is in the offline muon reconstruction (5.7\%) which includes an uncertainty (1\% per muon) to cover variations of the scale factor as a function of $\pt$ and $\eta$ of muons. Other systematic uncertainties include the uncertainty in the trigger (1.5\%), dimuon isolation (negligible), dimuon reconstruction effects related to overlaps of muon trajectories in the tracker and in the muon system (3.5\%), and dimuon mass shape, which affects the efficiency of the requirement that the two dimuon masses are compatible (1.5\%). The uncertainty in the LHC integrated luminosity of the data sample (2.2\%) is also included~\cite{CMS-PAS-SMP-12-008}. All the uncertainties quoted above, which relate to the final analysis selection efficiency for signal events, sum up to 7.4\%. The uncertainties related to the parton distribution functions (PDFs) and the knowledge of the strong coupling constant $\alpha_s$ are estimated by comparing the PDFs in CTEQ6.6~\cite{Nadolsky:2008zw} with those in NNPDF2.0~\cite{Ball:2010de} and MSTW2008~\cite{Martin:2009iq} following the PDF4LHC recommendations~\cite{Botje:2011sn}. Using the analysis benchmark samples, they are found to be 3\% for the signal acceptance. Varying the QCD renormalization/factorization scales has a negligible effect. The total systematic uncertainty in the signal acceptance and selection efficiency is 8.0\%.

\section{Results \label{sec:results}}
When the data satisfying all analysis selections were unblinded, no events were observed in the signal diagonal region, as illustrated in Fig.~\ref{fig:control}~(right). The expected background in the diagonal signal region is $1.0 \pm 0.5$ events, where the uncertainty accounts for both statistical and systematic effects. This background includes contributions from $\bbbar$ production and direct $\JPsi$ pair production, as discussed in Sec.~\ref{sec:background}.

\begin{figure*}[t]
\begin{center}
\includegraphics[width=0.49\textwidth]{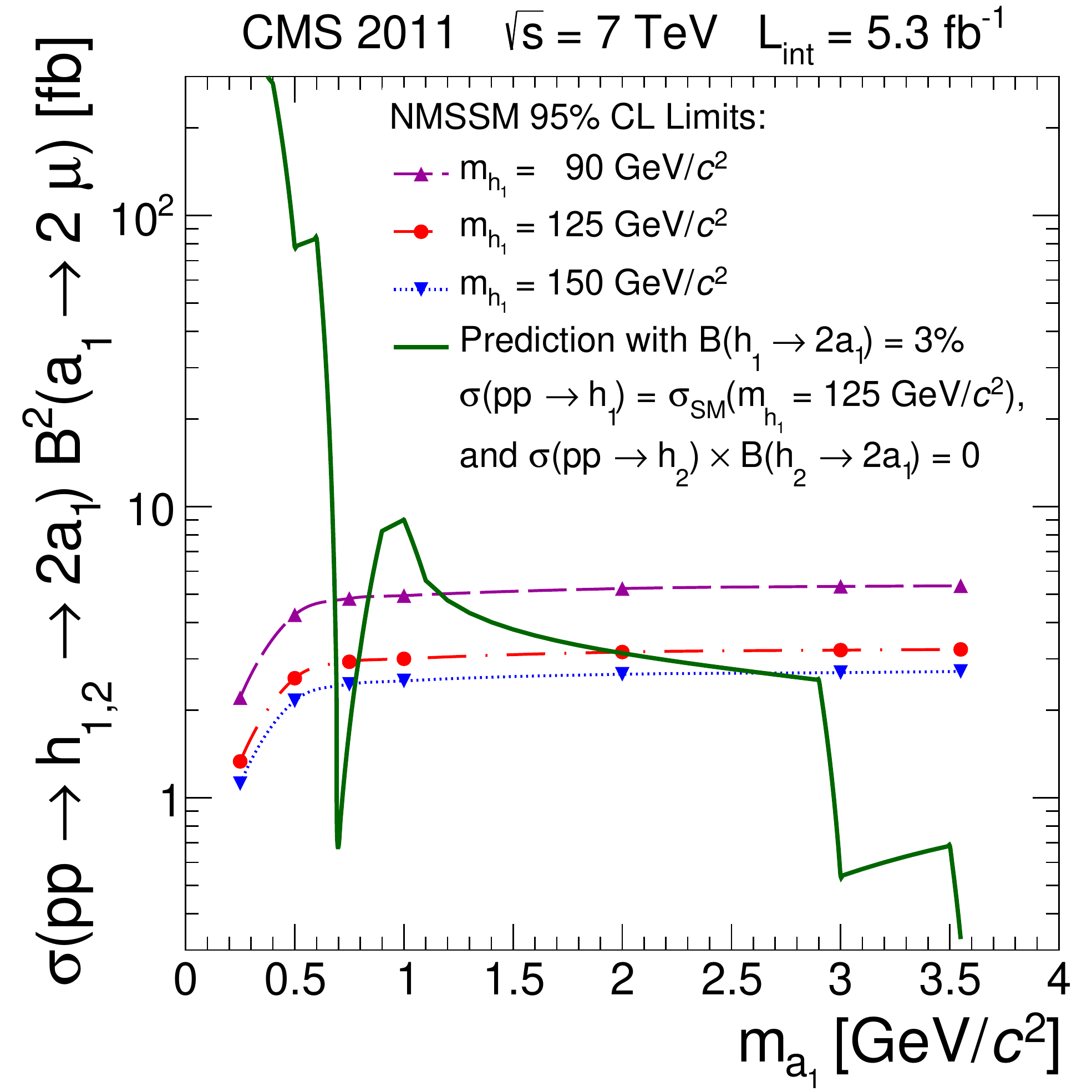}
\hfill
\includegraphics[width=0.49\textwidth]{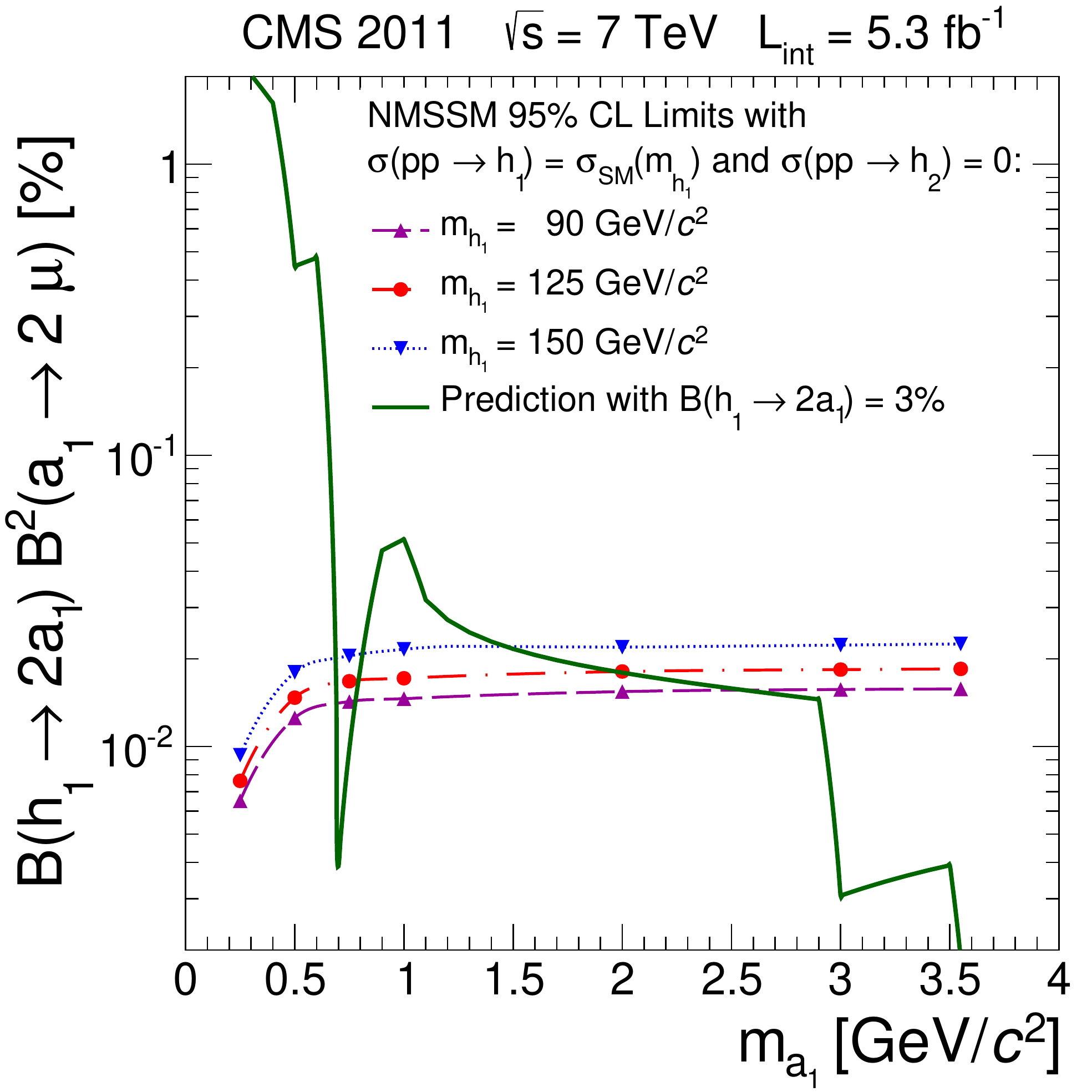}
\end{center}
\caption{Left: The $95\%$ CL upper limits as functions of $m_{\text{a}_1}$, for the NMSSM case, on $\sigma(\Pp\Pp \to \text{h}_{1,2} \to 2 \text{a}_1) \times \mathcal{B}^2(\text{a}_1 \to 2 \mu)$ with $m_{\text{h}_1}=90\GeVcc$ (dashed curve), $m_{\text{h}_1}=125\GeVcc$ (dash-dotted curve) and $m_{\text{h}_1}=150\GeVcc$ (dotted curve). The limits are compared to the predicted rate (solid curve) obtained using a simplified scenario with $\mathcal{B}(\text{h}_{1} \to 2 \text{a}_1) = 3\%$, $\sigma (\Pp\Pp \to \text{h}_{1})=\sigma_\mathrm{SM}(m_{\text{h}_{1}} = 125\GeVcc)$~\cite{Dittmaier:2011ti}, $\sigma (\Pp\Pp \to \text{h}_2) \times \mathcal{B}(\text{h}_{2} \rightarrow 2 \text{a}_1) = 0$, and $\mathcal{B}(\text{a}_1 \to 2\mu)$ as a function of $m_{\text{a}_1}$ which is taken from~\cite{Dermisek:2010mg} for NMSSM parameter $\tan \beta = 20$. Right: The $95\%$ CL upper limits on $\mathcal{B}(\text{h}_{1} \to 2 \text{a}_1) \times \mathcal{B}^2(\text{a}_1 \to 2 \mu)$ with $m_{\text{h}_1}=90\GeVcc$ (dashed curve), $m_{\text{h}_1}=125\GeVcc$ (dash-dotted curve) and $m_{\text{h}_1}=150\GeVcc$ (dotted curve) assuming $\sigma(\Pp\Pp \to \text{h}_{1}) = \sigma_\mathrm{SM}(m_{\text{h}_{1}})$~\cite{Dittmaier:2011ti} and $\sigma (\Pp\Pp \to \text{h}_2) \times \mathcal{B}(\text{h}_{2} \rightarrow 2 \text{a}_1) = 0$. The limits are compared to the predicted branching fraction (solid line) obtained using a simplified scenario with $\mathcal{B}(\text{h}_{1} \to 2 \text{a}_1) = 3\%$ and $\mathcal{B}(\text{a}_1 \to 2\mu)$ as a function of $m_{\text{a}_1}$ which is taken from~\cite{Dermisek:2010mg} for NMSSM parameter $\tan \beta = 20$.
\label{fig:results_a}}
\end{figure*}

For an arbitrary new physics model predicting the signature investigated in this Letter, the results can be presented as the 95\% confidence level (CL) upper limit:
\[ \sigma(\Pp\Pp \to 2 \text{a} + \text{X}) \times \mathcal{B}^2 (\text{a} \rightarrow 2\mu) \times \alpha_{\text{gen}} < 0.86 \pm 0.06\fbinv, \]
where $\alpha_{\text{gen}}$ is the generator-level kinematic and geometric acceptance defined in Sec.~\ref{sec:selection}. The calculation uses the value of the integrated luminosity $\mathcal{L} = 5.3\fbinv$ of the data, and takes the ratio $\epsilon_{\text{full}} / \alpha_{\text{gen}} = 0.67 \pm 0.05$, derived in Sec.~\ref{sec:selection}. This ratio includes the scale factor that corrects for experimental effects not accounted for by the simulation. The variation in this ratio over all of the used benchmark points is covered by systematic uncertainties. The limit is applicable to models with two pairs of muons coming from light bosons of the same type with a mass in range $0.25 < m_{\text{a}} < 3.55\GeVcc$ where the new light bosons are typically isolated, spatially separated to not be vetoed by the isolation requirement and have no substantial lifetime. The efficiency of the selections in this analysis abruptly deteriorates if the light boson's decay vertex is more than $\sim 4$~cm from the beamline in the transverse plane.

We interpret these results in the context of the NMSSM and the dark-SUSY benchmark models, taking into account the dependence of the signal selection efficiencies on $m_{\text{h}}$ and $m_{\text{a}}$ (see Tab.~\ref{tab:efficiency_NMSSM} and Tab.~\ref{tab:efficiency_SUSY}), and derive $95$\% CL upper limits on the product of the cross section and branching fraction, using a Bayesian prescription. We also compare the derived experimental limits with a few simplified prediction scenarios. In the representative models, for any fixed combinations of $m_{\text{h}}$ and $m_{\text{a}}$ both the Higgs boson production cross section and the branching fractions can vary significantly, depending on the choice of parameters. In the absence of broadly accepted benchmark scenarios, we normalize the production cross sections in these examples to that of the SM Higgs boson~\cite{Dittmaier:2011ti}.

For the NMSSM, the $95$\% CL upper limit is derived for $\sigma \left( \Pp\Pp \rightarrow \text{h}_{1,2} \rightarrow 2 \text{a}_1 \right) \times \mathcal{B}^2(\text{a}_1 \rightarrow 2\mu)$ as a function of $m_{\text{h}_1}$ for three choices of $m_{\text{a}_1}$ as shown in Fig.~\ref{fig:results}~(left) and as a function of $m_{\text{a}_1}$ for three choices of $m_{\text{h}_1}$ as shown in Fig.~\ref{fig:results_a}~(left). As $m_{\text{h}_2}$ is unrestricted for any given $m_{\text{h}_1}$, we use $\epsilon_{\text{full}}(m_{\text{h}_2}) = \epsilon_{\text{full}}(m_{\text{h}_1})$ to simplify the interpretation. This is conservative since $\epsilon_{\text{full}}(m_{\text{h}_2})>\epsilon_{\text{full}}(m_{\text{h}_1})$ if $m_{\text{h}_2}>m_{\text{h}_1}$, for any $m_{\text{a}_1}$. We also derive the $95$\% CL upper limit for $\mathcal{B}\left( \text{h}_{1} \rightarrow 2 \text{a}_1 \right) \times \mathcal{B}^2(\text{a}_1 \rightarrow 2\mu)$ as a function of $m_{\text{a}_1}$ for three choices of $m_{\text{h}_1}$ as shown in Fig.~\ref{fig:results_a}~(right) assuming that only $\text{h}_{1}$ gives a significant contribution to the final state considered in this analysis and has the production cross section of a SM Higgs boson, i.e. $\sigma(\Pp\Pp \to \text{h}_{1})  = \sigma_\mathrm{SM}(m_{\text{h}_{1}})$ and $\sigma(\Pp\Pp \to \text{h}_{2}) \times \mathcal{B}(\text{h}_{2} \rightarrow 2 \text{a}_1) = 0$. For the NMSSM simplified prediction scenario we use $\mathcal{B}(\text{a}_1 \rightarrow 2\mu)$ as a function of $m_{\text{a}_1}$, calculated in~\cite{Dermisek:2010mg} for $\tan{\beta}=20$ with no hadronization effects included in the $m_{\text{a}_1} < 2m_{\tau}$ region. The branching fraction $\mathcal{B}(\text{a}_1 \rightarrow 2\mu)$ is influenced by the $\text{a}_1 \rightarrow s \bar{s}$ and $\text{a}_1 \rightarrow gg$ channels. The significant structures in the predicted curves visible in Fig.~\ref{fig:results_a} arise from the fact that $\mathcal{B}(\text{a}_1 \rightarrow gg)$ varies rapidly in that region of $m_{\text{a}_1}$. The rapid variation in $\mathcal{B}(\text{a}_1 \rightarrow gg)$ occurs when $m_{\text{a}_1}$ crosses the internal quark loop thresholds. The representative value of $\mathcal{B}(\text{a}_1 \rightarrow 2\mu)$ is equal to $7.7\%$ for $m_{\text{a}_1} \approx 2\GeVcc$. Finally, we choose $\mathcal{B}(\text{h}_{1} \rightarrow 2 \text{a}_1) = 3\%$, which yields predictions for the rates of dimuon pair events comparable to the obtained experimental limits.

In the case of the dark-SUSY model, the $95$\% CL upper limit is derived for $\sigma(\Pp\Pp \rightarrow \text{h} \rightarrow 2 \text{n}_1 \rightarrow 2 \text{n}_D + 2\gamma_D) \times \mathcal{B}^2(\gamma_D \rightarrow 2\mu)$ as a function of $m_{\text{h}}$. This limit is shown in Fig.~\ref{fig:results}~(right) for $m_{\text{n}_1}=10\GeVcc$, $m_{\text{n}_D}=1\GeVcc$ and $m_{\gamma_D}=0.4\GeVcc$. For the dark-SUSY simplified prediction scenario we use the branching fraction $\mathcal{B}(\gamma_D \rightarrow 2\mu)$ close to its maximum at $m_{\gamma_D}=0.4\GeVcc$, of $45\%$, calculated in~\cite{Falkowski:2010cm}. We also use $\mathcal{B}(\text{n}_1 \rightarrow \text{n}_D + \gamma_D) = 50\%$, allowing for other possible decays. Finally, we choose $\mathcal{B}(\text{h} \rightarrow 2 \text{n}_1) = 1\%$, which yields predictions for the rates of dimuon pair events comparable to the obtained experimental limits.

The sensitivity of this search can be compared to that of a similar analysis performed at the Tevatron~\cite{Abazov:2009yi} after rescaling with the ratio of the Higgs boson cross sections at the LHC and the Tevatron. If plotted in Fig.~\ref{fig:results} and Fig.~\ref{fig:results_a}~(left), the Tevatron results would have exclusion limits above $\sim 130$\unit{fb}, therefore the search presented in this Letter has one order of magnitude better sensitivity compared to previous experimental constraints.

\section{Summary \label{sec:summary}}
A search for non-standard-model Higgs boson decays to pairs of new light bosons, which subsequently decay to pairs of oppositely charged muons ($\text{h} \to 2\text{a} + \text{X} \to 4 \mu + \text{X}$) has been presented. The search is based on a data sample corresponding to an integrated luminosity of $5.3\fbinv$ collected by the CMS experiment in proton-proton collisions at $\sqrt{s} = 7$\TeV in 2011. No excess is observed with respect to the SM predictions. An upper limit at $95\%$ confidence level on the product of the cross section times branching fraction times acceptance is obtained. The limit is valid for new light-boson masses in the range $0.25 < m_{\text{a}} < 3.55\GeVcc$ and for Higgs boson masses in the range $m_{\text{h}} > 86\GeVcc$. Although the results have been interpreted in the context of the NMSSM and the dark-SUSY benchmark models for $m_{\text{h}} < 150\GeVcc$, it is possible to extend them by smoothly extrapolating the model-independent cross section limit to higher masses. The analysis has been designed as a quasi-model-independent search allowing interpretation of its results in the context of a broad range of new physics scenarios predicting the same type of signature. In the context of the NMSSM and one of the SUSY models with hidden valleys this search provides the best experimental limits to date, significantly surpassing the sensitivity of similar searches performed at the Tevatron.

\section{Acknowledgments \label{sec:acknowledgments}}
We would like to thank Joshua Ruderman (LBNL and University of California at Berkeley) for his guidance with the theoretically motivated benchmark samples of dark SUSY and useful discussions. We wish to congratulate our colleagues in the CERN accelerator departments for the excellent performance of the LHC machine. We thank the technical and administrative staff at CERN and other CMS institutes, and acknowledge support from: FMSR (Austria); FNRS and FWO (Belgium); CNPq, CAPES, FAPERJ, and FAPESP (Brazil); MES (Bulgaria); CERN; CAS, MoST, and NSFC (China); COLCIENCIAS (Colombia); MSES (Croatia); RPF (Cyprus); Academy of Sciences and NICPB (Estonia); Academy of Finland, MEC, and HIP (Finland); CEA and CNRS/IN2P3 (France); BMBF, DFG, and HGF (Germany); GSRT (Greece); OTKA and NKTH (Hungary); DAE and DST (India); IPM (Iran); SFI (Ireland); INFN (Italy); NRF and WCU (Korea); LAS (Lithuania); CINVESTAV, CONACYT, SEP, and UASLP-FAI (Mexico); MSI (New Zealand); PAEC (Pakistan); SCSR (Poland); FCT (Portugal); JINR (Armenia, Belarus, Georgia, Ukraine, Uzbekistan); MST and MAE (Russia); MSTD (Serbia); MICINN and CPAN (Spain); Swiss Funding Agencies (Switzerland); NSC (Taipei); TUBITAK and TAEK (Turkey); STFC (United Kingdom); DOE and NSF (USA).

\bibliography{auto_generated}   
\cleardoublepage \appendix\section{The CMS Collaboration \label{app:collab}}\begin{sloppypar}\hyphenpenalty=5000\widowpenalty=500\clubpenalty=5000\textbf{Yerevan Physics Institute,  Yerevan,  Armenia}\\*[0pt]
S.~Chatrchyan, V.~Khachatryan, A.M.~Sirunyan, A.~Tumasyan
\vskip\cmsinstskip
\textbf{Institut f\"{u}r Hochenergiephysik der OeAW,  Wien,  Austria}\\*[0pt]
W.~Adam, E.~Aguilo, T.~Bergauer, M.~Dragicevic, J.~Er\"{o}, C.~Fabjan\cmsAuthorMark{1}, M.~Friedl, R.~Fr\"{u}hwirth\cmsAuthorMark{1}, V.M.~Ghete, J.~Hammer, N.~H\"{o}rmann, J.~Hrubec, M.~Jeitler\cmsAuthorMark{1}, W.~Kiesenhofer, V.~Kn\"{u}nz, M.~Krammer\cmsAuthorMark{1}, I.~Kr\"{a}tschmer, D.~Liko, I.~Mikulec, M.~Pernicka$^{\textrm{\dag}}$, B.~Rahbaran, C.~Rohringer, H.~Rohringer, R.~Sch\"{o}fbeck, J.~Strauss, A.~Taurok, W.~Waltenberger, G.~Walzel, E.~Widl, C.-E.~Wulz\cmsAuthorMark{1}
% ...
\vskip\cmsinstskip
\dag:~Deceased\\
1:~~Also at Vienna University of Technology, Vienna, Austria\\
% ...

\end{sloppypar}
\end{document}